# CCD Readout Electronics for the Subaru Prime Focus Spectrograph.

Stephen C. Hope[a], James E. Gunn[b], Craig P. Loomis[b], Roger E. Fitzgerald[c], Grant O. Peacock[a]

[a]Johns Hopkins University, Department of Physics and Astronomy, 3701 San Martin Drive, Baltimore, MD 21218, USA
[b]Princeton University, Department of Astrophysical Sciences, 4 Ivy Lane, Peyton Hall, Princeton, NJ 08544, USA
[c]Fitzgerald Engineering, 18006 Marshall Mill Road, Hampstead, MD 21074, USA

## ABSTRACT

The following paper details the design for the CCD readout electronics for the Subaru Telescope Prime Focus Spectrograph (PFS). PFS is designed to gather spectra from 2394 objects simultaneously, covering wavelengths that extend from 380 nm to 1260 nm. The spectrograph is comprised of four identical spectrograph modules, each collecting roughly 600 spectra. The spectrograph modules provide simultaneous wavelength coverage over the entire band through the use of three separate optical channels: blue, red, and near infrared (NIR). A camera in each channel images the multi-object spectra onto a 4k x 4k, 15 um pixel, detector format. The two visible cameras use a pair of Hamamatsu 2k x 4k CCDs with readout provided by custom electronics, while the NIR camera uses a single Teledyne HgCdTe 4k x 4k detector and Teledyne's ASIC Sidecar to read the device.

The CCD readout system is a custom design comprised of three electrical subsystems – the Back End Electronics (BEE), the Front End Electronics (FEE), and a Pre-amplifier. The BEE is an off-the-shelf PC104 computer, with an auxiliary Xilinx FPGA module. The computer serves as the main interface to the Subaru messaging hub and controls other peripheral devices associated with the camera, while the FPGA is used to generate the necessary clocks and transfer image data from the CCDs. The FEE board sets clock biases, substrate bias, and CDS offsets. It also monitors bias voltages, offset voltages, power rail voltage, substrate voltage and CCD temperature. The board translates LVDS clock signals to biased clocks and returns digitized analog data via LVDS. Monitoring and control messages are sent from the BEE to the FEE using a standard serial interface. The Pre-amplifier board resides behind the detectors and acts as an interface to the two Hamamatsu CCD's. The Pre-amplifier passes clocks and biases to the CCD's, and analog CCD data is buffered and amplified prior to being returned to the FEE.

In this paper we describe the detailed design of the PFS CCD readout electronics and discuss current status of the design, preliminary performance, and proposed enhancements.

# 1. INTRODUCTION

The Subaru Prime Focus Spectrograph is a spectroscopic system consisting of four identical bench-mounted spectrographs each with three optical channels covering the wavelength ranges 380-640nm, 640-955nm, and 955-1260nm simultaneously.

The spectrographs are fed with a 2394-actuator fiber positioning system at the prime focus of the Subaru telescope through a fiber cable to the fixed spectrographs; each spectrograph accepts 597 or 600 fibers. The near-IR channels use Teledyne H4RG 4K x 4K 15 micron-pixel HgCdTe detectors, and the shorter (blue and red) channels side-butted Hamamatsu fully-depleted 2K x 4K 15-micron pixel devices to make a 4K x 4K focal plane, with geometry essentially identical to the IR channel. The IR detectors will use the Teledyne SIDECAR ASIC for control and data acquisition, controlled by the SAM interface card. We have developed a new CCD control system for the blue and red CCD detectors, and this paper describes this system.

The detectors are typical of modern fully depleted CCDs, much like the LBL devices developed for NASA's SNAP program[1]. The CCDs for the red channel are 200 microns thick and have a coating that yields 90% quantum efficiency at about 800 nm. The blue CCDs are electronically identical but are thinned to 100 microns and have a blue-enhanced coating which yields roughly 90 percent QE at 450 nm.

Each chip has 4 serial registers, so the readout for each is 512 x 2048 (actually 512 x 2116) pixels. The serial registers terminate in a standard follower/reset MOSFET pair, and there is a p-channel JFET follower mounted to the CCD package for each output. The sensitivity of the output transistor with its summing node and the JFET follower is about 4uV per electron, and the read noise at our output rate of 75kpx/s is 3 electrons. Full pixel well is about 150,000 electrons; we will digitize with about 1.5 electrons/ADU, so a 16-bit ADC saturates at about 100,000 electrons and we resolve about half a standard deviation in the noise from the output FET.

The control electronics must generate the rather complex clocking signals for the chips, drive the chip with them, accept the eight channels of analog video from the two-chip mosaic, process it by standard (analog) CDS circuitry, digitize it, demultiplex the digital data into an image, cache the image locally, add metadata, and transfer it to an external server.

In addition, temperature monitoring and control data are handled passively by the system and handed to an external temperature controller.

The desired performance for the control system include a noise contribution of less than 0.7 electrons from the electronics themselves, the ability to bin arbitrarily in the vertical direction (which is the dispersion direction in the spectrographs), and to implement all of the various modes necessary to prepare the devices for integrating and reading.

The visible camera electronics consist of a number of subsystems with architecture similar to Hyper Suprime-Cam[2].

The system in each visible camera consists of:

- Two Hamamatsu 2K x 4K S10892-1628 (blue) or S10892-1629 (red) fully depleted p-channel CCDs in a coplanar 4K x 4K array
- A custom preamplifier featuring 8 channels of low-noise amplification provided by JFET-input operational amplifiers, each with a gain of six. The preamps also provide DC restoration circuitry. This preamp is in the vacuum.
- Custom front-end electronics (FEE), which perform CDS processing on the video outputs, digitizes these signals and outputs the digital data serially over LVDS, and generates the clock waveforms for the CCD from input LVDS timing signals and supplied DC rail voltages. The FEE is housed on an interface card part of which is in the vacuum and part outside.
- An off-the shelf PC104-based Linux Intel x86 computer as back-end electronics (BEE) in which most of the real-time signal processing is performed in a commercial FPGA daughter card. This unit communicates with a remote server both for control and data transmission via gigabit Ethernet.

The system is encapsulated in the block diagram below:

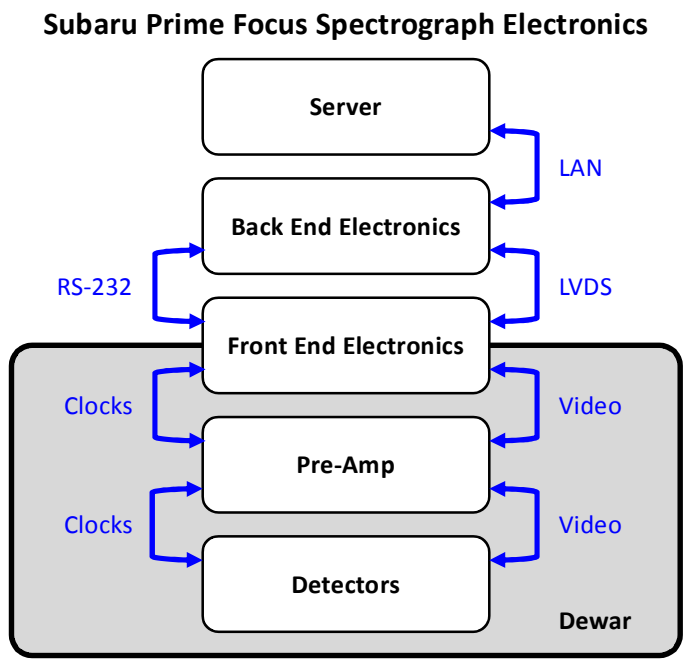

Figure 1, System Block Diagram

The paper discusses each of the elements of the system in turn, preliminary performance and proposed improvements.

# 2. CCD ARRAY/PREAMPLIFIER

The Preamplifier board is used to interface control and clock signals to a coplanar pair of 2k x 4k Hamamatsu CCD's. Figure 2 details the mechanical assembly comprising of the Preamplifier and a pair of CCD's. The board is located in the optical path, immediately behind the CCD array. Connection between the preamp and each CCD is made with a flexible printed circuit board (FPC). On one end a custom 59 pin socket array mates with the CCD. On the other, a 65 pin AirBorn Nano connector interfaces with the Preamplifier. The FPC uses two layers of copper tracks and is designed to readily bend to follow the path from the CCD.

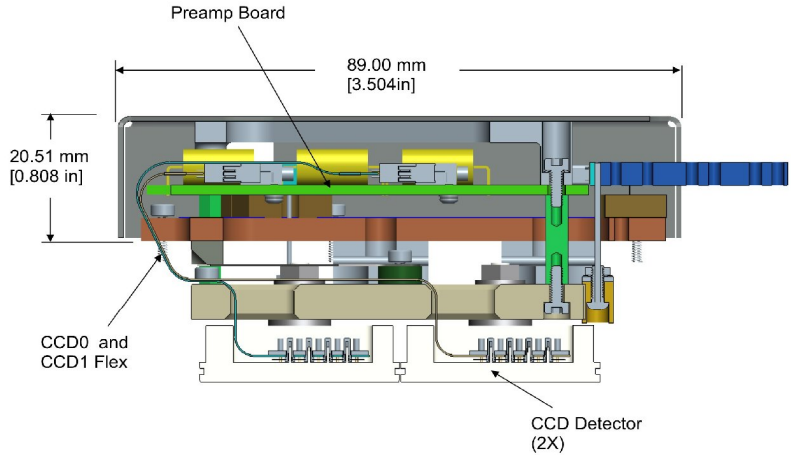

Figure 2, Preamp/Detector Assembly. The preamplifier board in located in the optical path immediately behind the coplanar CCD array.

Each CCD has four (4) output channels, hence the preamplifier comprises of eight (8) low noise amplifiers. Each utilizes an Analog Devices AD8610 in an MSOP8 surface mount package. The CCD output is approximately 5μV per photon. Each amplifier is configured with a gain of six (6), which roughly equates to 1½ photons per ADU, with a full-scale range of 100,000 photons.

To minimize common mode noise, analog signals between the Preamplifier and the front-end electronics (FEE) are driven as pseudo-balanced output pairs. Each pair comprises of: A signal line with 47.5Ω resistor in series with the AD8610's output impedance, resulting in a nominal 50 Ω impedance; and a return line connected to the amplifiers analog ground pin via matching 50 Ω series resistance.

In addition to buffering the analog signal, the preamplifier board also serves as a conduit, routing various clock and bias signals from the Front End Electronics (FEE) to the CCD.

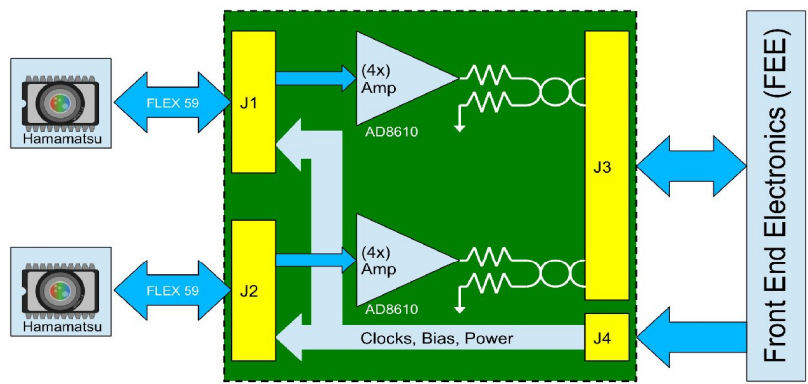

Figure 3. Preamplifier functional block diagram.

# 3. FRONT END ELECTRONICS (FEE)

The Front End Electronics (FEE), as detailed by figure 4, is a custom board that integrates the following functionality:

- CCD clock conditioning/biasing
- Double correlated sampling (CDS)
- Analog to digital conversion (ADC)
- Power supply monitoring and control
- Clock bias voltage monitoring
- CCD temperature monitoring

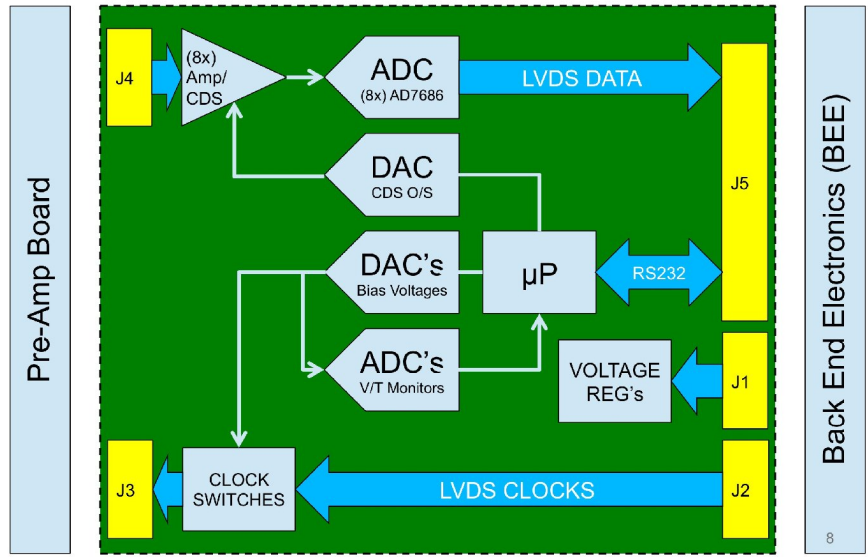

Figure 4. Functional block diagram of the FEE board.

Figure 5 shows a prototype of the FEE used in lab testing.

**3.1 CCD Clock Conditioning**

The Back End Electronics (BEE) manages clock generation. Clock signals are routed from the BEE to the FEE using low voltage differential signaling (LVDS). Texas Instruments LVDT390's convert each LVDS clock signal to CMOS (TTL) levels, each of which drive a single pole double throw solid-state switch. The normally open (NO) and normally closed (NC) terminals are biased with voltages supplied by an Analog Devices AD5361, 16 channel, 16 bit DAC. This results in a clock signal with fully programmable positive and negative swing. A passive, single pole filter provides slew rate limiting.

Despite the fact that the clock biases for each CCD will likely be the same, the design includes provision for biasing each CCD separately.

**3.2 Correlated Double Sampling (CDS)**

Each analog CCD output channel is buffered a P-channel JFET follower integral to the CCD. Prior to clocking charge into the buffer, the buffer is reset. The value of the buffer at reset is uncertain as some charge remains. This uncertainty comes from kTC noise, or reset noise. To negate this error the pedestal voltage (i.e. the voltage after reset) is subtracted from the signal of interest in a process known as correlated double sampling. The FEE utilizes analog CDS comprising of two switched inputs. The integrator is first reset to a known state by shorting the integrating capacitor. The pedestal voltage is then subtracted (integrate minus) by integrating through the inverting channel. The signal of interest is then added (integrate plus) by integrating the non-inverting input. The resultant output is the difference with noise and offset (i.e. pedestal) removed.

The CDS circuit is subject to other potential noise sources. These include: Dielectric adsorption error in the integrator capacitor; integrator gain error resulting from clock jitter; amplifier offset error; and integrator output noise integral to the amplifier. To minimize dielectric adsorption error the integrator uses a 1nF polypropylene capacitor. To minimize clock jitter gain errors all clock signals are generated by an FPGA integral to the backend electronics (BEE). DAC programmable offsets allow amplifier offset to be removed. Finally to minimize amplifier noise, the output signal is filtered at the input of the ADC.

### 3.3 Analog to Digital Conversion (ADC)

CDS signals are converted from analog to digital by eight (8) Analog Devices AD7686, 16-bit analog to digital converters – one per analog channel. Converters are grouped into two groups of 4 – one per CCD. Data from each group is serialized and clocked out 50MHz via LVDS.

### 3.4 Power Supplies & Backplane Biasing

The FEE utilizes multiple supplies for analog and digital control. On-board, low dropout regulators with high pwer supply rejection ratio (PSRR) regulate each supply. PSSR is further enhanced by an additional capacitor-inductor-capacitor Pi filter at the input. Each supply can be enabled or disabled under microprocessor control. In addition, the microprocessor provides supply voltage monitoring.

The CCD back bias voltage (Vbb) is configurable via an Analog Devices AD5361, 16-bit DAC driving an OPA545 amplifier with active low pass filtering. The amplifier features a microprocessor controlled enable pin, providing a secondary means of disabling Vbb.

### 3.5 Clock Bias Monitoring

In addition to monitoring each power supply the micro also facilitates clock bias voltage monitoring. Each channel is set and then read back to ensure the proper voltage was applied.

### 3.6 CCD Temperature Monitoring

Each CCD includes a 1k-Ohm thermistor for temperature monitor. The FEE utilizes a Texas Instruments ADS1247 to monitor temperature. The ADS1247 is a fully integrated device featuring two precision, programmable current sources, and an on-board ADC.

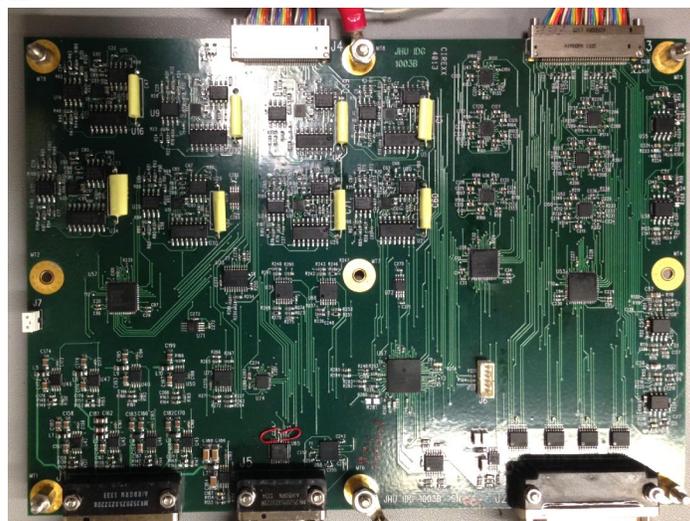

Figure 5. Prototype front-end electronics (FEE) board.

## 4. BACK END ELECTRONICS (BEE)

The back end electronics (BEE) comprises of two off-the-shelf, PCIe/104 circuit boards manufactured by RTD Embedded Technologies Inc. (www.rtd.com) – the CMX32MCS1200-2048 embedded PC with the FPGA35S6045HR field programmable IO board. The primary function of the BEE is to manage image capture. This includes configuring peripherals, gathering telemetry, formatting images, transferring images, and generating CCD, CDS and DAC clocks.

### 4.1 CMX32MCS1200-2048 Embedded PC

Figure 6 is an image of the CMX32MCS1200-2048 Embedded PC (CMX32). The CMX32 is a powerful PC/104 single board computer with a PCIe/104 stackable bus structure, which includes the following features:

- Intel Celeron M (ULV 722) processor
- 2GB memory
- Eight x1 PCIe links
- One x16 PCIe link (configurable as one x8 or one x4)
- 4GB onboard industrial SATA flash disk
- Two SATA ports
- Four Serial Ports (RS-232/422/485)
- Six USB 2.0 Ports
- Dual Gigabit Ethernet
- High Definition Audio
- Analog VGA & LVDS Panel
- Advanced Digital I/O (aDIO)
- Advanced Analog I/O (aAIO)
- Stackable PCIe bus
- -40 to +85°C standard operating temperature

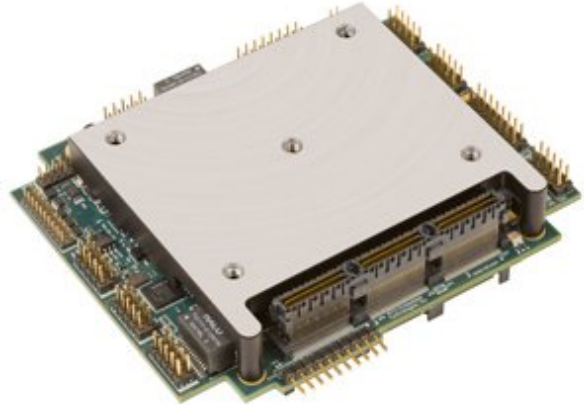

Figure 3, CMX32MCS1200 Embedded PC

The CMX32 manages all non-timing-critical (asynchronous) events. In regard to CCD readout, this includes: FPGA and FEE configuration; FITS image formatting; and image upload. Other functionality includes managing other system peripherals, gathering telemetry and communicating with the Subaru messaging hub. Software for the CMX32 is written using a combination of C and Python running in a Debian Linux Embedded Operating System.

### 4.2 FPGA35S6045HR

Figure 7 is an image of the FPGA35S6045HR (FPGA) daughter card. The FPGA is a field programmable gate array module designed to interface with the CMX32 using the PCIe bus. It includes the following features:

- PC/104 form factor
- PCIe/104 stackable bus structure
- Xilinx Spartan-6 FPGA
- XC6SLX45T
- 46,661 Logic Cells
- 2,489 Kb of internal RAM
- 48, 5V tolerant Digital I/O
    - Selectable pull-up/pull-down per byte
    - Pull-up can be 3.3V or 5V
    - ESD Protected
    - Can be used as LVDS IO or LVTTL IO
- 40 3.3V tolerant High-Speed I/O
    - Can be used as LVDS Input or LVTTL IO

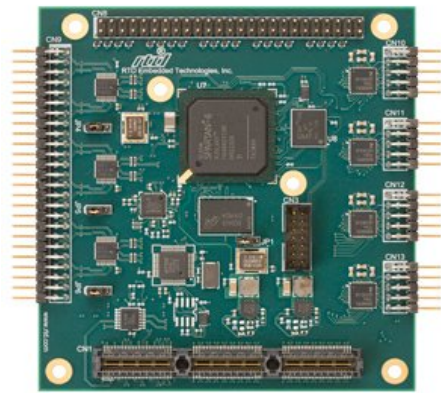

Figure 4, FPGA35S6045HR

The FPGA manages all timing-critical (synchronous) events. This includes generating the CCD, CDS and ADC clocks, as well as transferring image data to CMX32 accessible RAM. The FPGA interfaces with the CMX32 via a PCIe bus. Interface with the FEE is via LVDS (low voltage, differential signaling), which is integral to the Xilinx Spartan 6 architecture.

Functionality of the FPGA is defined using VHDL. Figure 8 describes the program architecture. In this instance, the FPGA is configured as a 'waveform processor unit' or WPU. The WPU is analogous to a CPU in that it executes instructions fetched from memory. However, in contrast to a CPU, timing is extremely precise. It is designed to generate clocks with sub-nanosecond precision and a granularity of 80ns. WPU memory comprises of 128KB of block RAM accessible to both the CMX32 and the FPGA.

WPU binaries are created by software running on the CMX32. Each instruction defines the state of each CCD/CDS clock lines at that point in time, and specifies the number of cycles to dwell before executing the next instruction. A single bit is used to trigger an ADC read which comprises of 65 clock pulses streamed at a rate of 50MHz. Clearly, this affords exceptional flexibility, allowing the definition of routines to read, wipe, erase and so on. A typical binary will include all the necessary clock transitions to read a single row of pixels, and a loop count that defines the number of rows to read.

The FPGA also includes the functionality to synchronize operation of up to 8 cameras. In a multiple camera application, one BEE is defined as a master, the rest as slaves. The WPU engine's execution is synchronized to a 25MHz clock generated by the master BEE's FPGA. This clock is routed to the master WPU and up to 7 slave WPU's across equal length LVDS lines. Execution starts when the clock begins and is synchronous to the clock, consequently WPU execution is synchronous across all units.

The FPGA module also includes 128MB of DDR2 RAM. This is made accessible to both the WPU and the CMX32. This is used as temporary repository for image data streamed by the WPU, where, upon completion, it is subsequently retrieved by the CMX32. A raw, 4k x 4k image requires approximately 36MB's of memory, hence this is sufficient.

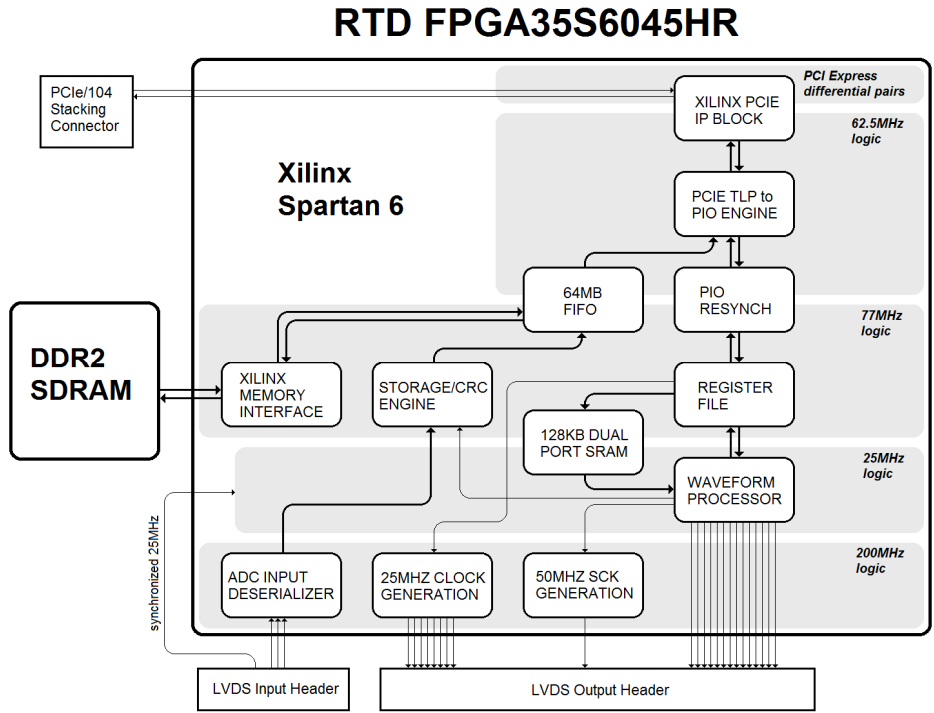

Figure 8. FPGA architecture. The pictorial describes the programmed architecture of the FPGA, which is used to generate clocks, transfer and compile images, and synchronize camera operation.

# 5. PRELIMINARY PERFORMANCE

To date, performance testing has been conducted using a pair of 'fake' CCD's (figure 9) coupled by flex cables to the preamplifier board. The 'fake' CCD's include passive circuitry that, through clock mixing, allows the generation of a pseudo video signal. In addition, they include provision for terminating or grounding the outputs, and/or injecting custom waveforms via an external arbitrary function generator.

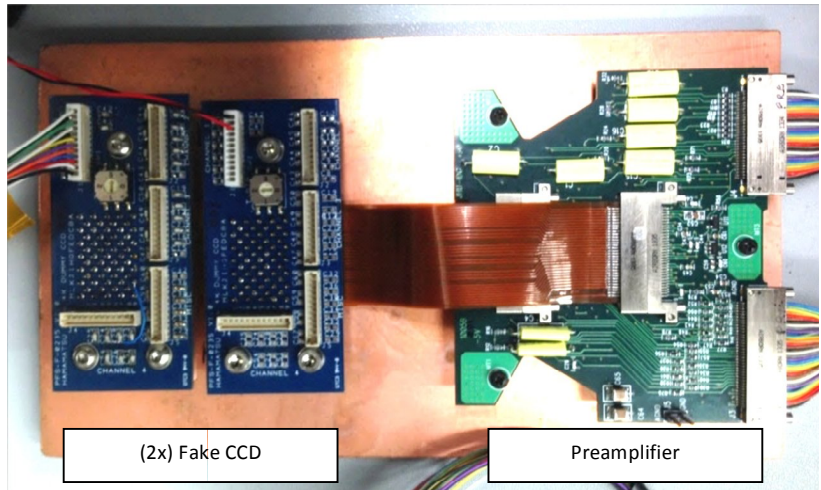

Figure 9. Preamplifier test setup. The image depicts the preamplifier on the right, attached to a pair of fake CCD's on the left, via a pair of rigid-flex cable assemblies.

## 5.1 First 'Fake' Light

The first 'fake' light image detailed by figure 10 was generated using a pair a fake CCDs. One was configured to provide four channels of video (4 grey scale stripes), the other had four channels terminated to ground with 100-Ohm resistors.

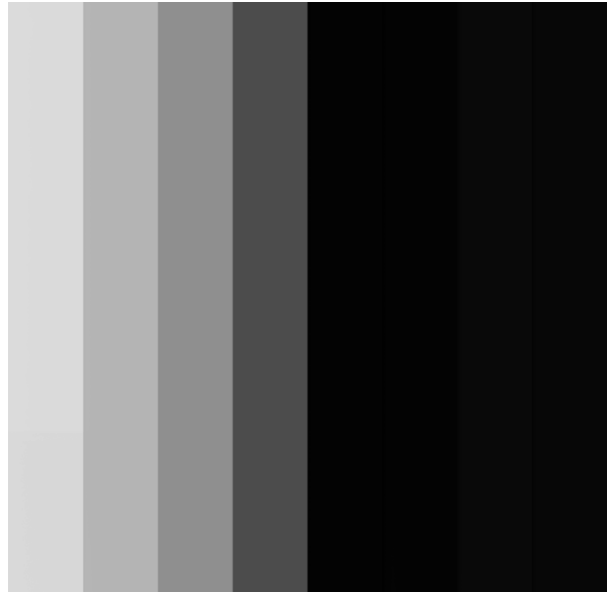

Figure 10. First 'fake' light. This image was generated using a pair of fake CCD's. The four gray scale stripes on the left were created by a pseudo-video signal generated by mixing the clocks. Each stripe represents one channel on the first CCD. The black area on the right represents the data for the four channels on the second fake CCD, which were grounded via 100-Ohm resistors.

## 5.2 Readout performance

Each CCD has four output channels, hence a total of eight channels for the 4k x 4k coplanar array, which are read synchronously. A channel comprises a parallel-in, serial-out, shift register. A single row read comprises of a parallel load of one row of pixels followed by 536 serial shifts (512 active pixels, plus overscan). To read the entire array, this process is repeated 4240 times, again including overscans. A parallel load takes 240μs, while each serial shift takes 13.4μs, hence, time required to read a row approximately 7.4ms, which roughly equates to 31.5s for the entire array.

The system architecture affords a slower transfer rate for the purpose of testing and debugging. However, to date all tests have been performed at the target clock rate without error.

The next revision of the FEE as detailed in section 6.1 will increase ADC data requirements from 16-bits per pixel to 18. Despite the fact that this represents a 12.5% increase in serial data, there will be no impact on readout performance as the additional time required to transfer data will compensated for by a minor reduction in integration time.

## 5.3 Noise floor

To assess noise floor the CCD channel 0 input was tied to ground with a 100-Ohm termination resistor. A DAC offset was applied to the analog CDS. 512 consecutive samples were captured at full data rate. Analysis was performed to assess the standard deviation. Figure 11 details the resultant waveform, which had a standard deviation of 0.79, marginally higher than the targeted deviation of 0.5.

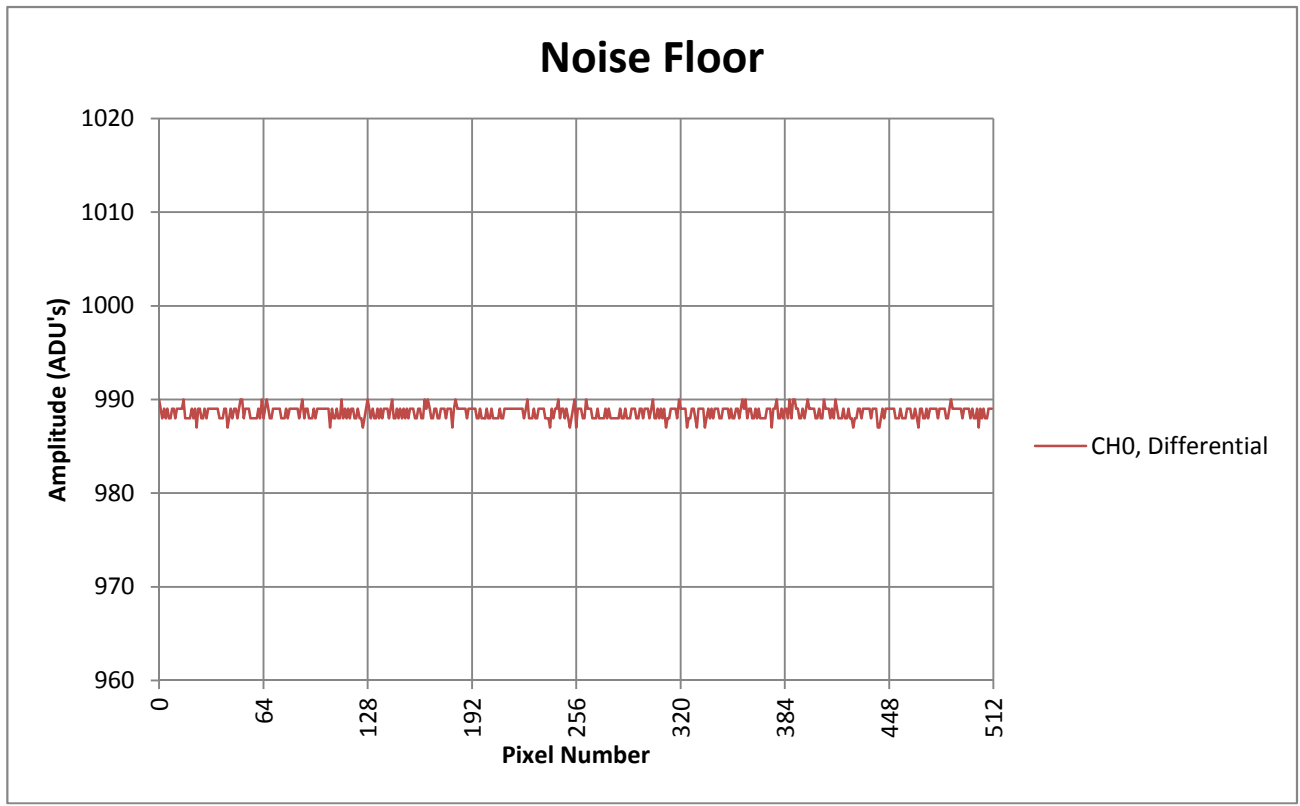

Figure 11, Noise floor. Noise floor was measured with the input to the preamplifier grounded via 100-Ohm resistor. A DAC offset was applied to the CDS.

### 5.4 Common mode noise rejection

To assess common mode noise rejection, CDS channel 0 was configured with a fully differential input, while channel 1 was configured with a single-ended input, both with unity gain. Each was driven by an individual pre-amplifier channel, terminated with a 100-Ohm resistor. An offset was applied to the integrator. An external noise source was placed in close proximity to the video lines between preamplifier and the BEE. The results are depicted in figure 12. With the noise source present, the differential channel had a standard deviation of 0.93, approximately 0.15 greater than the nominal noise floor. In contrast the single-ended channel was almost twenty times greater with deviation of 17.31.

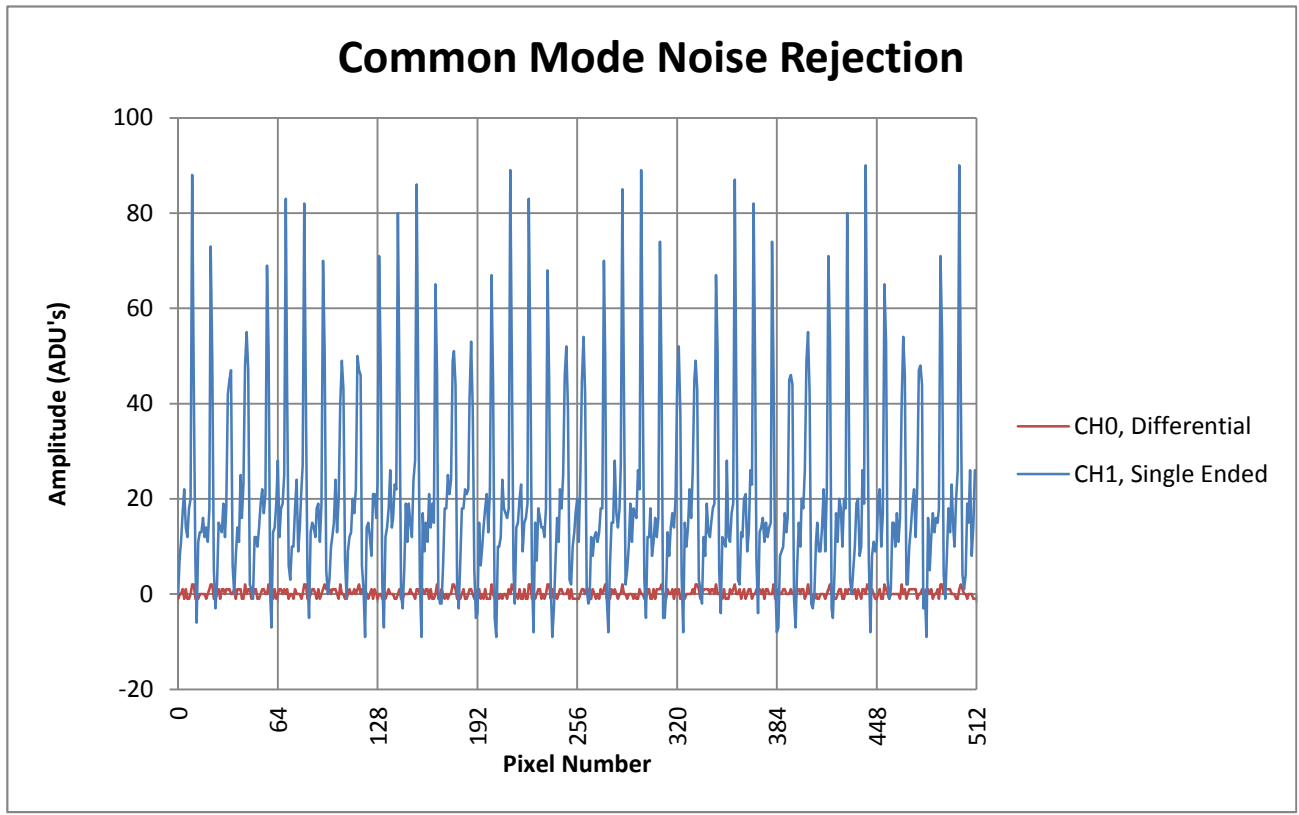

Figure 12. Common Mode Noise Rejection. Common mode noise was measured with the input to each channels preamplifer grounded via a 100-Ohm resistor. A DAC offset was applied to each channel. A noise source was placed in proximity to the analog wiring between the preamplifier and the FEE. Channel 0 was setup as a differential input, channel 1 as a single ended input.

# 6. PATH FORWARD

In the current configuration, efforts to minimize noise have resulted in a noise floor with a 0.8 standard deviation. The target for noise floor is 0.5. Proposed enhancements include:

- Increasing ADC resolution to 17 bits
- Enhancing FEE board layout to reduce power supply induced noise

In addition, further testing will include crosstalk between adjacent channels, DNL and INL error measurements, and clock jitter analysis.

## 6.1 Increased ADC Resolution.

Input referred noise or code transition noise is common source of ADC noise[3]. It is internal to the ADC and is a function of resistor and kTC noise. This is illustrated in Figure 13. An ideal ADC would have zero transition noise and code transition would be extremely precise (figure 13a). In reality code output less determinate, with transition noise often upwards of one LSB (figure 13b).

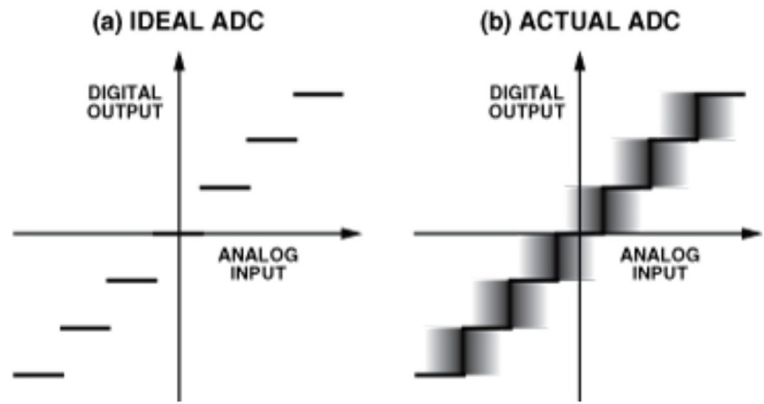

Figure 13, ADC input referred noise. Image (a) shows the performance of an ideal ADC, where code transitions are precise and predictable. Image (b) shows typical performance where code transitions are less deterministic.

The prototype FEE incorporates an AD7686 single-ended, 16-bit converter, with 0.5 LSB transition noise. This will be replaced with an AD7690, 18-bit differential converter. In a single-ended application, one MSB is lost, reducing the resolution to 17 bits. In addition, one LSB will be dropped by the BEE. This will result in the following 16-bit performance gains:

- Transition noise, reduced from 0.5LSB to 0.38LSB (nominal)
- Differential non-linearity, reduced from 0.5LSB to 0.25LSB (nominal)
- Integral non-linearity, reduced from 0.6LSB to 0.38LSB (nominal)

## 6.2 Enhanced FEE Board Layout

The present iteration of the board has two ground planes and multiple power planes, each on different layers. In some instances power and ground planes overlap. This allows noise to couple from one plane to another. Moreover, ideally the analog and digital grounds should meet at the ADC. The FEE has 8 ADC's each digitizing a CCD channel, each of which should have an individual ground. The next iteration of the FEE will strictly adhere to these requirements.

The original prototype, as detailed by figure 5, was supposed to reside inside the cryostat requiring a flange with hermetic feed-through connectors to bridge the cryostat wall. In the new design the board will act as a hermetic feed-through. Signals will be routed from inside to outside the cryostat on inner layers. Annulus rings on the outer layers will provide a mating surface for O-ring seals. This is similar to a design implemented by Caltech for the Zwicky Transient Facility Mosaic[3].

A proof-of-concept board was built assess leak rates. Figure 14 shows the proof-of-concept board both on the bench and installed on the test chamber. Helium leak testing was conducted using an RGA from Scientific Research Systems. The results showed no detectable leaks.

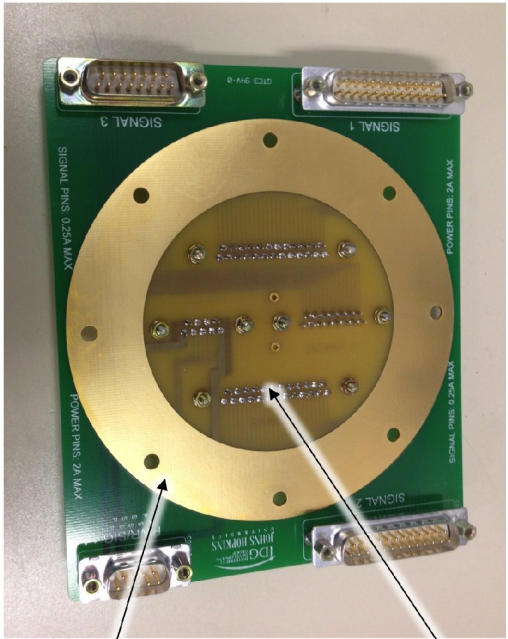 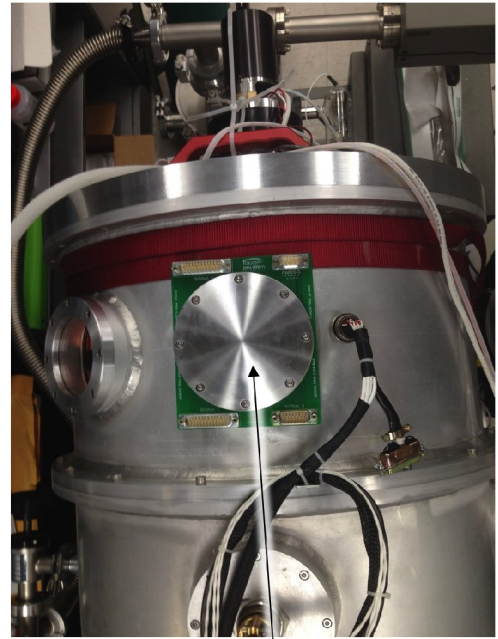

Figure 14, Proof-of-concept hermetic feed through PCB. The left hand image shows the board with gold plated annulus ring, common to both outer layers, which act as O-ring mating surfaces. The right hand image shows the board installed on a cryostat.

The new design, shown conceptually in figure 15, has many benefits:

- The FEE will be accessible from the outside, facilitating repair or replacement without disassembling the

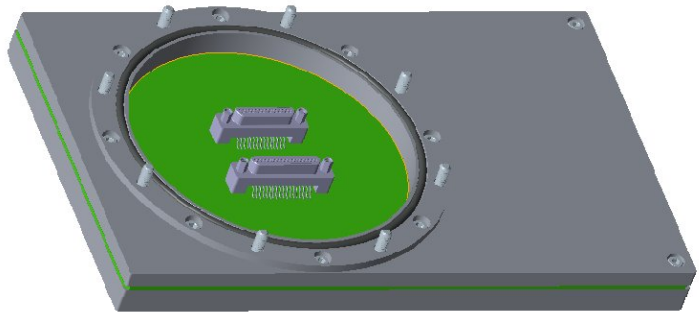

- cryostat.
- It will negate the need for a hermetic feed-through connectors and associated cabling, reducing both complexity and cost.
- It will allow quiet analog circuits to reside inside the cryostat, while simultaneously allowing noisy digital circuits to remain outside.

Figure 15.CAD rendition of the new FEE board. The board is mounted in housing which is designed to mount with an O-ring seal to the outside of the camera cryostat.

## ACKNOWLEDGMENTS

We gratefully acknowledge support from the Funding Program for World-Leading Innovative R&D in Science and Technology (FIRST), program: "Subaru Measurements of Images and Redshifts (SuMIRe)", CSTP, Japan